\documentclass{osa-article}
\usepackage{subcaption}
\usepackage[font=normalsize,labelfont={bf}]{caption}
\usepackage{hyperref}
\usepackage{float}
\usepackage{xcolor} %added by Frank to show edits

\journal{oe}
\begin{document}

\title{Electric Field Measurement of Femtosecond Time-Resolved Four-Wave Mixing Signals in Molecules}

\author{Francis Walz,\authormark{1} Siddhant Pandey,\authormark{1} Liang Z. Tan, \authormark{3} and Niranjan Shivaram \authormark{1,2,*}}

\address{\authormark{1} Department of Physics and Astronomy, Purdue University, West Lafayette, Indiana 47907, USA\\
\authormark{2} Purdue Quantum Science and Engineering Institute, Purdue University, West Lafayette, Indiana 47907, USA\\
\authormark{3} Molecular Foundry, Lawrence Berkeley National Laboratory, Berkeley, California 94720, USA}

\email{\authormark{*}niranjan@purdue.edu} %% email address is required

% \homepage{http:...} %% author's URL, if desired

%%%%%%%%%%%%%%%%%%% abstract %%%%%%%%%%%%%%%%
\begin{abstract}
We report an experiment to measure the femtosecond electric field of the signal emitted from an optical third-order nonlinear interaction in carbon dioxide molecules. Using degenerate four-wave mixing with femtosecond near infrared laser pulses in combination with the ultra-weak femtosecond pulse measurement technique of TADPOLE, we measure the nonlinear signal electric field in the time domain at different time delays between the interacting pulses. The chirp extracted from the temporal phase of the emitted nonlinear signal is found to sensitively depend on the electronic and rotational contributions to the nonlinear response. While the rotational contribution results in a nonlinear signal chirp close to the chirp of the input pulses, the electronic contribution results in a significantly higher chirp which changes with time delay. Our work demonstrates that electric field-resolved nonlinear spectroscopy offers detailed information on nonlinear interactions at ultrafast time scales. 
\end{abstract}

%In addition to being sensitive to the mechanisms generating the nonlinear signal, this approach can allow measurement of dephasing time in the nonlinear interaction.    

%%%%%%%%%%%%%%%%%%%%%%%%%%  body  %%%%%%%%%%%%%%%%%%%%%%%%%%
\section{Introduction}
 In an all optical measurement of light-matter interactions, the amplitude and phase of the emitted electric field provide complete information. With electric field metrology demonstrated at THz - PHz frequencies \cite{keiber2016, sederberg2020, ridente2022, liu2022}, field-resolved spectroscopy at high sensitivity is now possible in this frequency range \cite{pupeza2020}. In the case of nonlinear optical spectroscopy such as four-wave mixing spectroscopy, the emitted signal is related to the induced third-order polarization in the medium and is determined by the third-order nonlinear response tensor $\chi^{(3)}_{ijkl}$ that can offer insight into the electronic character of the system being studied. Measuring the electric field of the nonlinear signal on ultrafast time scales offers access to the real-time behavior of the nonlinear polarization and hence the real and imaginary parts of the $\chi^{(3)}_{ijkl}$ tensor, as the system dynamically evolves. 
 
 Typical nonlinear spectroscopy techniques, like optical Kerr effect (OKE) spectroscopy \cite{wieman1976,lotshaw1987} and degenerate four-wave mixing (DFWM) \cite{shirley1980}, measure the intensity of the nonlinear signal light, which is proportional to the magnitude of $\chi^{(3)}_{ijkl}$. In the optical heterodyne (OHD) configuration of OKE, a local oscillator is introduced which allows a measurement of the real and imaginary parts of $\chi^{(3)}_{ijkl}$ \cite{mcmorrow1988, palese1994}. One drawback of such OHD techniques is the requirement of multiple measurements in different local oscillator configurations. Spectral domain techniques like Single Shot Supercontinuum Spectral Interferometry (SSSI) \cite{kim2002, chen2008} simplify measurement by using chirped supercontinuum probe pulses where frequency is mapped to time delay after the pump pulse. In these experiments, the phase shift acquired by the probe pulse at different time delays after the pump pulse in gas-phase targets is measured in a single-shot. Other experiments have measured the modification of ultrashort pulses due to propagation in solid nonlinear media by completely measuring the ultrashort pulse \cite{rivet2000, piredda2005}. Here, we use the spectral interferometry technique of TADPOLE \cite{fittinghoff1996}, which is capable of measuring ultra-weak, ultrashort pulses down to the zeptojoule level, to completely measure the real-time varying electric field of a DFWM nonlinear signal, as a function of time delay between the interacting pulses, in carbon dioxide molecules. We find that the chirp of the emitted nonlinear signal electric field sensitively depends on the electronic and rotational contributions to the nonlinear signal.      
 
 %d imaginary parts of $\chi^{(3)}$, optical heterodyning with a weak local oscillator is needed. Other techniques like transient absorption spectroscopy (TAS) that use resonant probe light, operate in a regime where the imaginary part of $\chi^{(3)}$ is dominant. Another approach is interferometry, in which both the phase and amplitude of the signal can be measured. With interferometry, we can measure real and imaginary parts of the signal without the need for heterodyning. Examples include Single Shot Supercontinuum Spectral Interferometry (SSSI) \cite{chen2008}, and methods such as those found in \cite{rivet2000} and \cite{piredda2005} that measure small changes in the refractive index of a material. Similar direct electric field measurements have been done with attosecond pulses as in \cite{sederberg2020}. Since the emitted electric field contains the complete imprint of the involved nonlinear process, a direct measurement can provide additional information.  By using the technique of TADPOLE (Temporal Analysis by Dispersing a Pair Of Light E fields), devised by Fittinghoff et. al.\cite{fittinghoff1996}, we measure the emitted electric field from ground state CO$_2$, in a DFWM experiment.
 
\section{Experimental Method}
\begin{figure}[b]
\centering\includegraphics[width= 0.99 \textwidth]{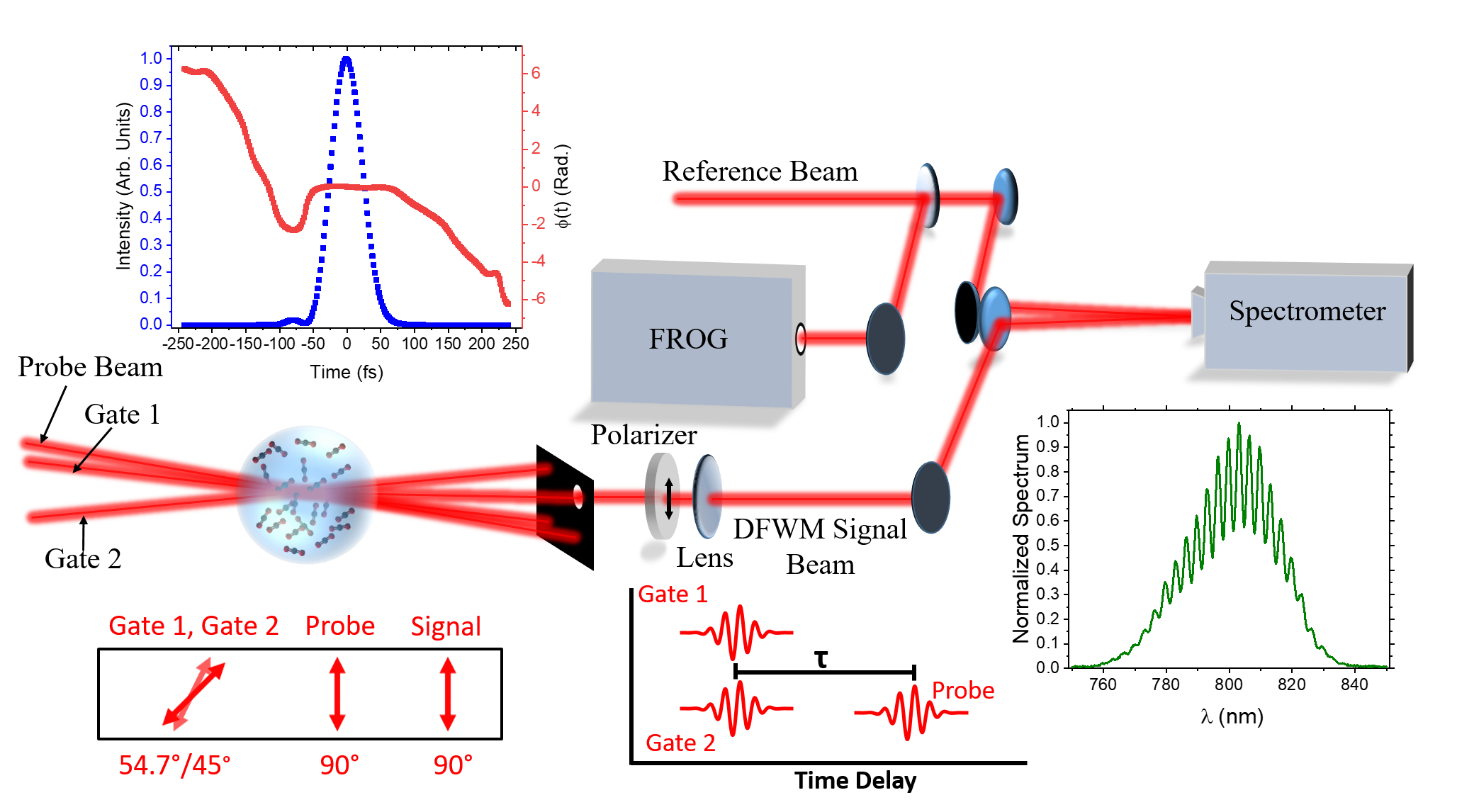}
\caption{Schematic of the TADPOLE setup for measuring the DFWM signal. The DFWM signal is isolated from the Gate and Probe beams using a mask and sent through a polarizer and lens into the TADPOLE measurement part of the setup. A Reference beam (completely characterized using a FROG apparatus) is combined with the weak DFWM signal pulse in a spectrometer where spectral interference fringes are measured. A TADPOLE reconstruction algorithm is then used to extract the amplitude and phase of the DFWM signal electric field. The top left inset shows the FROG measurement of the Reference pulse and the bottom right inset shows typical fringes due to spectral interference of the DFWM signal and the Reference pulse. The Gate, Probe and Signal polarizations are shown at the bottom left. The Gate pulses are phase-locked with each other and the Probe arrives late at positive time delays.}
\label{fig1:Expt}
\end{figure}

Near infrared pulses centered at 800 nm, with a pulse width of 50 fs and a repetition rate of 1 KHz are split to form Gate and Probe pulses. The Gate pulse is divided into two phase-locked Gate pulses (Gate 1 and Gate 2) using a mask. The Probe pulse is delayed with respect to the Gate pulses using an optical delay stage, and the Gate polarization can be rotated relative to the Probe polarization. A Reference pulse is derived from the Probe pulse for spectral interferometry. The Gate and Probe pulses are focused into a gas cell with a lens (figure \ref{fig1:Expt}). Spatial and temporal overlap are achieved using second harmonic generation in a beta barium borate (BBO) crystal. The gas cell is $\sim 90$ mm long, with 1 mm thick UV fused silica (UVFS) windows that do not distort the pulses significantly. The gas cell is filled with carbon dioxide gas at a pressure of $\sim$ 1 bar. After the interaction, the Gate and Probe beams are blocked by a second mask, allowing only the DFWM signal beam to pass. The Signal is passed through a polarizer set parallel to the Probe polarization. The Signal pulses generated in this experiment are in the pico-joule regime. A fully characterized Reference pulse is combined with the Signal pulse in a spectrometer to obtain the spectral interference measurement necessary for TADPOLE. Frequency Resolved Optical Gating (FROG) \cite{trebino1993} is used to completely characterize the Reference pulse.

The measured spectral interference of the DFWM signal and the Reference can be written in general as: 

\begin{equation}\label{eq:1}
\begin{split}
S(\omega) &= S_{R}(\omega) + S_{S}(\omega) + S_{B}(\omega)+\sqrt{S_{S}(\omega)}\sqrt{S_{B}(\omega)}\cos(\varphi_{S}(\omega)-\varphi_{B}(\omega)) \\ &+\sqrt{S_{R}(\omega)}\sqrt{S_{S}(\omega)}\cos(\varphi_{S}(\omega)-\varphi_{R}(\omega)+\omega\, \tau_{R}) \\ &+\sqrt{S_{R}(\omega)}\sqrt{S_{B}(\omega)}\cos(\varphi_{R}(\omega)-\varphi_{B}(\omega)+\omega\, \tau_{R})
\end{split}
\end{equation} 

$S(\omega)$ is the spectral intensity, $\omega$ is the angular frequency, and $\varphi(\omega)$ is the spectral phase. The subscripts $B$, $S$ and $R$ stand for background, signal and reference, respectively. $\tau_{R}$ is the time delay between the Reference and the Signal and the Reference and the background. It is noted that a constant background that is phase-coherent with the Signal light can interfere with the Signal electric field, and in general must be accounted for. In our experiment, we remove any such background by measuring DFWM in a BOXCARS geometry \cite{shirley1980} where the signal is emitted at a different angle and hence can be easily isolated, making this measurement background free. Thus, we can ignore $S_{B}(\omega)$ and $\varphi_{B}(\omega)$ in equation \ref{eq:1}, leading to the following simplified expression:

\begin{equation}\label{eq:2}
\begin{split}
S(\omega) &= S_{R}(\omega) + S_{S}(\omega) +\sqrt{S_{R}(\omega)}\sqrt{S_{S}(\omega)}\cos(\varphi_{S}(\omega)-\varphi_{R}(\omega)+\omega \,\tau_{R})
\end{split}
\end{equation}

To extract the Signal phase $\varphi_{S}(\omega)$, we Fourier transform $S(\omega)$ with respect to $\omega$. The first two terms on the right-hand side of equation \ref{eq:2} are related to the pulse spectrum, and are slowly varying. The last term contains the spectral interference fringes between the Reference and the Signal, and has fast oscillations. The Fourier transform spectrum, thus, has a distinct non-zero frequency peak corresponding to the fringe period (see Supplemental document).  We filter and shift this fast Fourier component to zero (to remove the $\omega\tau_{R}$ term), and then inverse Fourier transform the resulting spectrum \cite{takeda1982} from which we can extract $\varphi_{S}(\omega) - \varphi_{R}(\omega)$. Since the Reference phase $\varphi_{R}(\omega)$ is known we can extract $\varphi_{S}(\omega)$. We measure the Signal spectrum $S_{S}(\omega)$ for each Gate-Probe time delay by blocking the reference, which, along with the phase $\varphi_{S}(\omega)$ allows a complete measurement of the DFWM signal electric field ($E(\omega)$ or $E(t)$) at these different delays. Further, by carefully choosing the Gate-Probe relative polarization, we remove almost all the molecular rotation contributions to the nonlinear signal, and measure only the electronic contribution. This magic angle DFWM configuration \cite{lessing1976,siebert2000,reichert2015,ferdinandus2013,baskin1987} is discussed further in the next section.  

\section{Results and Discussion}
%\subsection{Reconstructed Signal Electric Field for CO$_{2}$}

\begin{figure}[t]
\centering
\includegraphics[width = 0.6 \textwidth]{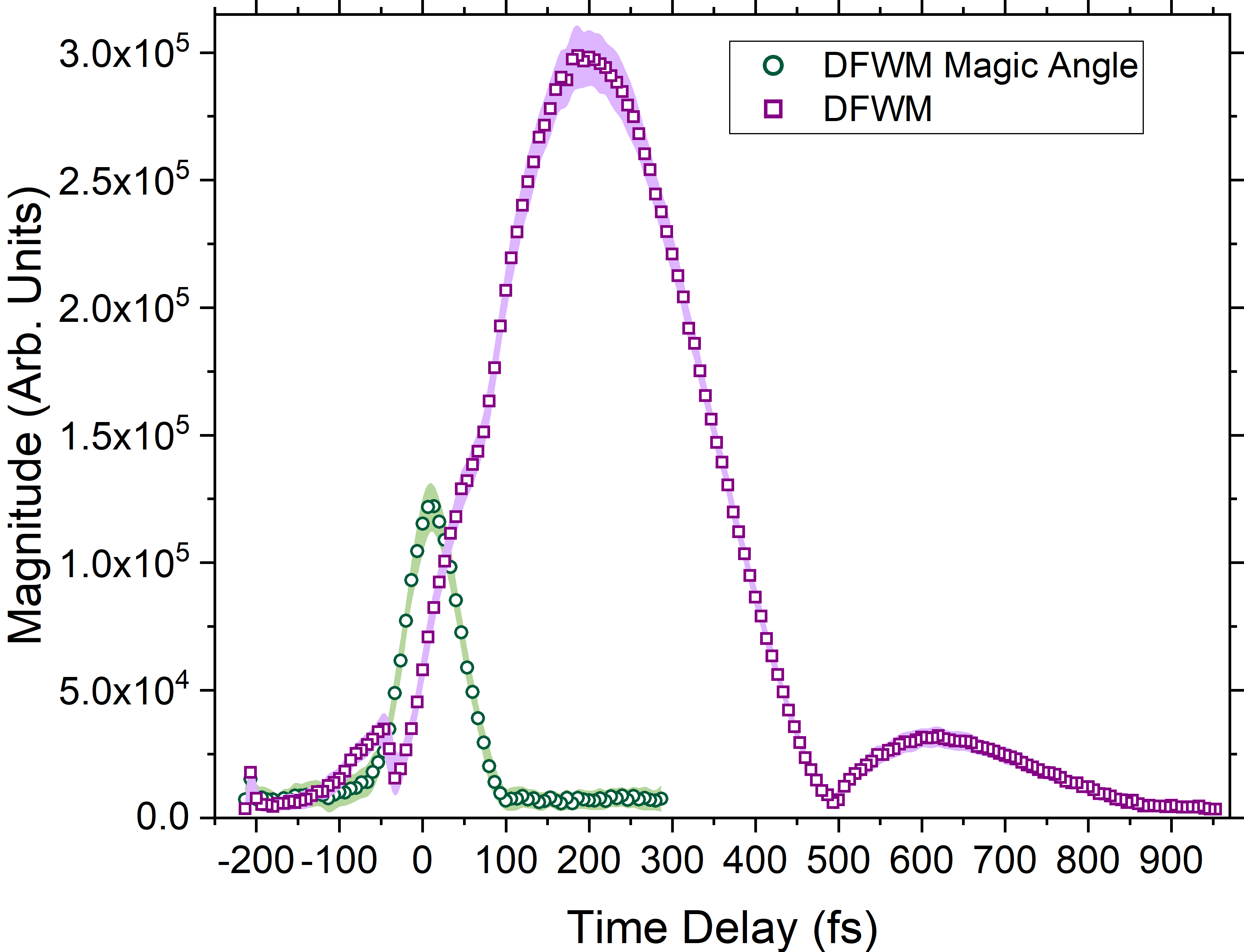}
\caption{Magnitude of the nonlinear signal obtained by integrating $E(\omega)$ over all frequencies, from CO$_{2}$ for DFWM (purple circles) and magic angle DFWM (green squares) schemes. The delayed rotational response peaked around 200 fs (seen in purple) is clearly absent in the magic angle configuration.}
\label{fig2:homodyne-magicangle}
\end{figure}

In a four-wave mixing interaction involving gas-phase molecules, the third-order nonlinear response has multiple contributions. A purely electronic contribution due to the laser field driven distortion of the electron cloud results in a nearly instantaneous response. A rotational contribution due to laser driven impulsive alignment of the molecule leads to both a near-instantaneous as well as a delayed response \cite{wahlstrand2013}. A vibrational contribution may exist provided the bandwidth of the laser pulses is sufficient to excite vibrational modes. In our experiment, the Gate pulses have an intensity of $\sim 5$ $TW/cm^2$ which is sufficient to impulsively align CO$_2$ molecules in the target. When the Probe and Gate pulses interact with the gas molecules, this results in a nonlinear optical signal containing both electronic and rotational contributions. We do not observe a signal corresponding to vibrational motion under the conditions of our experiment. Figure \ref{fig2:homodyne-magicangle} shows the frequency integrated third-order nonlinear signal magnitude as a function of time delay $\tau$ between the Gate and Probe pulses for the DFWM and magic angle DFWM schemes. In the DFWM scheme, the relative angle between the Gate and Probe pulse polarizations is set to 45 degrees whereas in the magic angle DFWM scheme this angle is set to 54.7 degrees. The magic angle configuration significantly suppresses the rotational contribution \cite{lessing1976,siebert2000,reichert2015,ferdinandus2013,baskin1987}. It is seen that the strong delayed response that peaks around a time delay of 200 fs in the DFWM scheme is absent in the magic angle DFWM scheme. Further, the signal near pulse overlap around zero time delay is also different in the magic angle case indicating that the rotational contribution in that region is suppressed as well. Complete removal of the rotational contribution in the pulse overlap region is possible only when the Gate and Probe pulses are of different frequencies, which is not the case in our experiment. However, we observe significant suppression of the rotational contribution to demonstrate clear differences in the properties of the emitted signal electric field between the electronic and rotational contributions to the nonlinear signal.    

\begin{figure}[h]
\centering\includegraphics[width= 0.99 \textwidth]{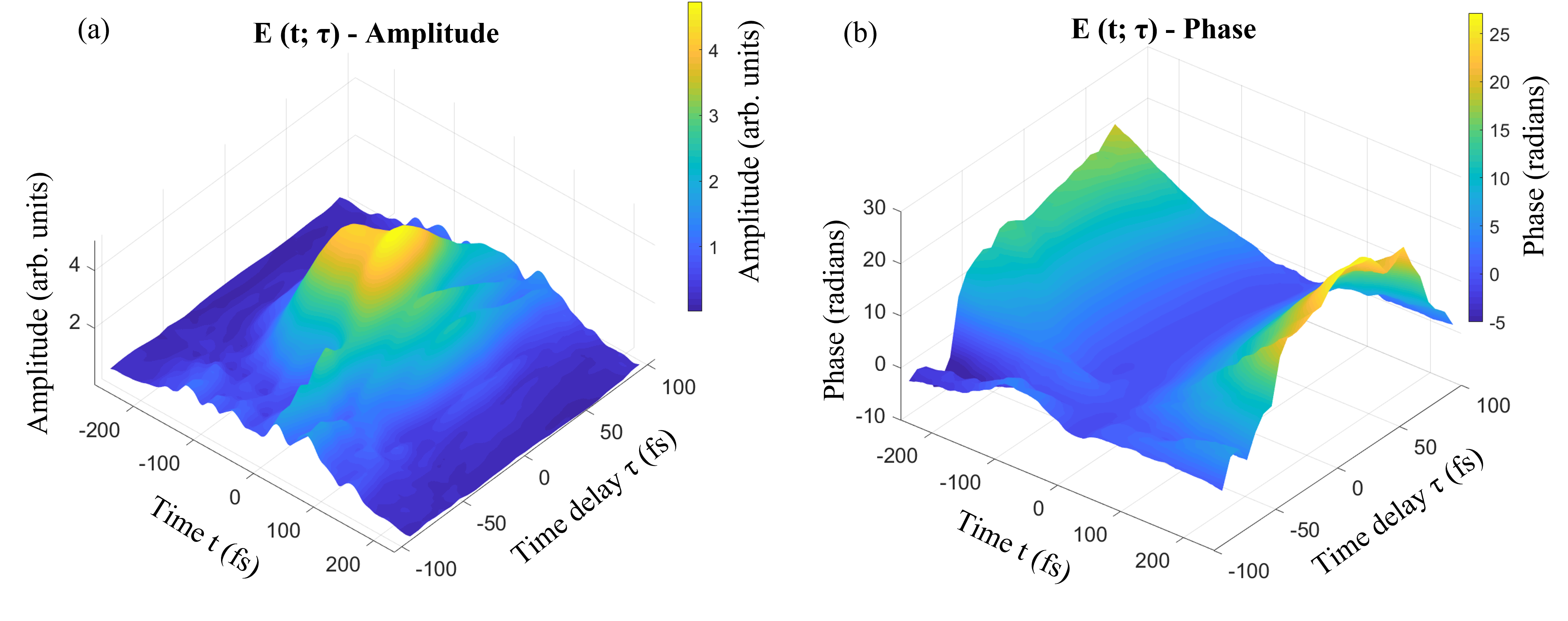}
\caption{Measured nonlinear signal electric field for the magic angle DFWM configuration. (a) Field amplitude as a function of time and time delay. (b) Field phase as a function of time and time delay.}
\label{fig3:E_t_timedelay}
\end{figure}

The time (or frequency) dependent nonlinear signal electric field amplitude and phase contains complete information available in a given nonlinear interaction. While figure \ref{fig2:homodyne-magicangle} contains time-resolved information about the nonlinear interaction which can be obtained from several different FWM techniques, our approach of measuring the nonlinear signal electric field can provide further insight into the nonlinear interaction, in real-time, as the time-dependent third-order polarization induced in the molecule coherently oscillates and generates the electric field. Using the measured spectrum and phase ($S_{S}(\omega)$, $\varphi_{S}(\omega)$), we obtain the total electric field $E(\omega)=\sqrt{S(\omega)}e^{-i\varphi(\omega)}$ which is Fourier transformed to obtain the time-domain electric field ${E}(t)= E_0(t)e^{-i\phi(t)}$ where $E_0(t)$ is the time-dependent amplitude and $\phi(t)$ is the time-dependent phase. Note that the carrier frequency-dependent term $e^{-i\omega_{o}t}$ is omitted for simplicity. Fig \ref{fig3:E_t_timedelay} shows the retrieved nonlinear signal electric field amplitude and phase for multiple Gate-Probe time delays $\tau$ as a function of time $t$, in the magic angle DFWM configuration (see Supplemental document for lineouts of the electric field amplitude and phase at different time delays).

% The suppression of the delayed rotational response in magic angle DFWM can be seen from Fig. \ref{fig3:E_t_timedelay}. The magic angle DFWM signal has it's maximum strength close to $\tau$= 0, whereas the DFWM signal has its maximum when $\tau$ = 220 fs (not shown in Fig. \ref{fig3:E_t_timedelay}).

%\subsection{Measurement and Calculation of Nonlinear Signal Electric Field }
The electric field measured as a function of time $t$ and time delay $\tau$ between the Probe and Gate fields can be written as:

\begin{equation}\label{eq:e-field}
\begin{split}
E(t;\tau) &= E_0(t;\tau)e^{-i \phi(t;\tau)}
\end{split}
\end{equation}

The measured phase $\phi(t;\tau)$ can be modeled as a polynomial function in $t$:
\begin{equation}\label{eq:phase}
\begin{split}
\phi(t;\tau) &= \phi_0(\tau) + a(\tau)\,t + b(\tau)\,t^{2}  + c(\tau)\,t^{3}  + d(\tau)\,t^{4}
\end{split}
\end{equation}

\begin{figure}[t]
\centering\includegraphics[width= 0.99 \textwidth] {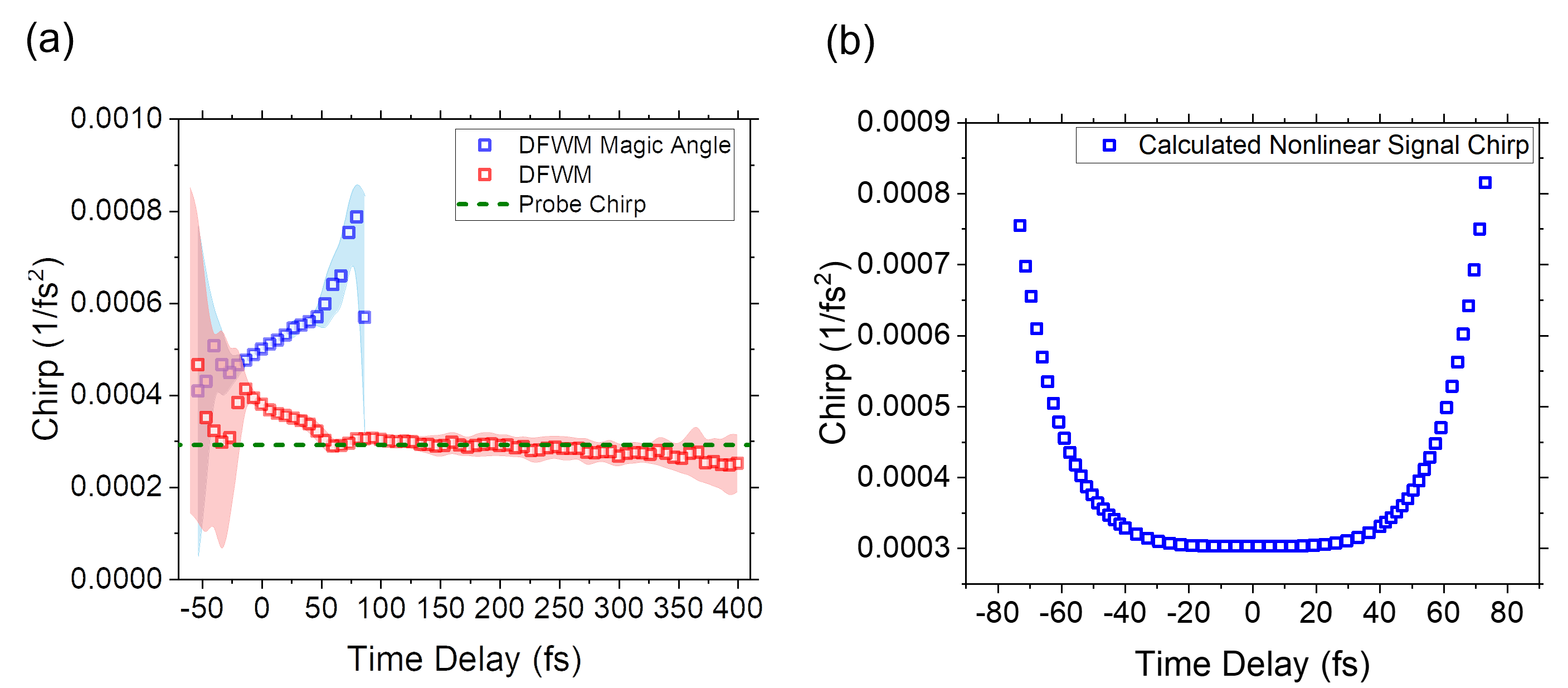}
\caption{(a) Chirp of the nonlinear signal electric field extracted from a polynomial fit of the time and time delay-dependent phase, for the case of DFWM and magic angle DFWM configurations. The shaded area represents error of one standard deviation. The chirp of the input Probe pulse is shown by the dashed line.  (b) Theoretical calculation of the nonlinear signal electric field chirp under the conditions of the experiment, for the case of an isolated carbon dioxide molecule with only the electronic contribution to the nonlinear signal.}
\label{fig4:bparam}
\end{figure}

Figure \ref{fig4:bparam} (a) shows the fit coefficient $b(\tau)$, which is the chirp of the electric field, as a function of time delay $\tau$ for the DFWM and magic angle DFWM cases. The $c(\tau)$ and $d(\tau)$ parameters are two orders of magnitude smaller than the $b(\tau)$ parameter and are thus neglected. In the DFWM case, the chirp increases near zero time delay where the Gate and Probe pulses overlap and then returns to the input chirp value for time delays beyond 75 fs. This indicates that the electronic contribution to the nonlinear signal (dominant near zero time delay) adds significant chirp to the nonlinear signal whereas the rotational contribution does not change the chirp compared to the chirp of the input pulses. The magic angle DFWM signal chirp increases rapidly from zero time delay until $\sim$ 75 fs where the signal level decreases significantly (see magic angle DFWM signal magnitude in figure \ref{fig2:homodyne-magicangle}) as no signal due to rotational contribution is present in this case. This further indicates that the time delay-dependent chirp in the nonlinear signal is electronic in origin. Plasma nonlinearities \cite{wahlstrand2013} are not expected to contribute at the intensity of $\sim$ 5 $TW/cm^2$ used in the experiment.        

%Additional information about the nonlinear process involved can be learned from the polynomial coefficients of different order. In Fig. \ref{fig4:bparam}, we plot the $b$ parameter as a function of gate-probe time delay.

%a) Quadratic chirp parameter $b$ (see Eq. \ref{eq:phase}) as a function of gate-probe time delay, for DFWM (red) and "magic angle" DFWM (blue) signals in gas-phase CO$_2$. b) Full Width at the half maximum of a Gaussian fit for the DFWM (red) and DFWM magic angle (blue) |E(t)| as seen in Fig. \ref{fig3:E_t_timedelay} as a function of time delay for gas-phase CO$_2$

% \begin{figure}[htbp]
% \centering\includegraphics[width= 0.8 \textwidth]{fig5_mag_chirp_ortho_analyzer.png}
% \label{fig5:chirp_amp_anlayzer_ortho}
% \caption{Magnitude of the measured E-field and chirp as a function of delay between the gate and probe pulses for the case of analyzer angle orthogonal to probe polarization at 2.5 TWcm$^{-2}$ intensity of the gate and probe. The electronic contribution is significantly suppressed as seen in the E-field magnitude near zero time delay whereas the chirp parameter shows large variations. As the rotational contribution increases and reaches a peak near a delay of 200 fs, the chirp decreases towards the value of the input probe chirp.}
% \end{figure}

%The $c$ and $d$ parameters were two orders of magnitude smaller than the $b$ parameter and can be neglected.
%\textcolor{red}{Theory to explain the significance of the chirp parameter.}

To investigate the origin of the chirp, we performed Lindblad equation simulations of the electronic response of a CO$_2$ molecule under the pulse sequence used in the experiments. We used a four level model of a CO$_2$ molecule in its bent room temperature geometry (bond angle $\sim$175 degrees) \cite{jensen2020}. The model consists of the ground state and the first three excited singlet states at 9.058 eV ($B_2$), 10.731 eV ($B_2$), and 12.916 eV ($B_2$). The results do not change upon inclusion of more states, suggesting that the chirp of the third-order response is dominated by these low-lying states. The model is parameterized by energy levels $\Omega$, and permanent and transition dipole moments ($\vec{\mu}$) between these states. These parameters were calculated using coupled-cluster singles and doubles (CCSD) method, using the Dalton software~\cite{aidas2014}. The evolution of the electronic state ($\rho$) under the pulse sequence is given by 
\begin{equation}\label{eq:lindblad}
\dot{\rho}(t) = -\frac{i}{\hbar}[H(t),\rho(t)] + \mathcal{L}_D \rho(t)
\end{equation}
with the Hamiltonian
\begin{equation}
H(t) = \Omega + \vec{\mu} \cdot (\vec{E}_1(t) + \vec{E}_2(t) + \vec{E}_3(t) )
\end{equation}

In these simulations, we have used pulses $\vec{E}_1(t)$,  $\vec{E}_2(t)$,  $\vec{E}_3(t)$ with durations, intensities, chirp, and polarizations that are the same as the experiment. Dephasing and population relaxation times of 300 fs were included via the Lindbladian $\mathcal{L}_D$, but were found to have an insignificant effect on the results as the signal is non-zero only during the duration of pulse overlap, at time scales much shorter than the dephasing and relaxation times. The result of integrating Eq.~\ref{eq:lindblad} is the time domain polarization $\vec{P}(t) = \mathrm{Tr}[\vec{\mu} \rho(t)]$. To extract the third-order nonlinear signal electric field, we perform third-order finite-difference derivatives of the polarization to calculate the signal as $\partial E_1 \partial E_2  \partial E_3 \vec{P}(\omega)$, for $\omega$ at the pulse frequency.  We obtain values of signal chirp that are enhanced from the chirp of the input pulses, increasing to a value of $b=0.0008\, \mathrm{fs}^{-2}$ at a time delay of 75 fs  (see figure \ref{fig4:bparam} (b)), which agrees well with our experimental measurement (figure \ref{fig4:bparam} (a)). These simulations support our interpretation that enhancement of the chirp observed in the experiment is electronic in origin. We attribute the asymmetric time delay-dependence of the chirp in the measurements to phase matching along the selected signal beam direction which restricts measurement to signal generated near zero time delay (pulse overlap) and positive time delay (Probe arriving late). The nonlinear signal generated when the Gate and Probe pulses exchange roles (negative time delay) is emitted along a different phase matching direction and not measured in our experiment, whereas the simulations contain the chirp of the entire third-order response, and hence, appears symmetric around zero time delay. The electric field chirp is, thus, a sensitive measure of the electronic contribution to the nonlinear signal even when the electronic contribution is dominated by the rotational contribution as seen in a homodyne measurement (figure \ref{fig2:homodyne-magicangle}). This demonstrates that electric field-resolved nonlinear spectroscopy could offer new insights into nonlinear interactions on ultrafast time scales.

%(b) The full-width at half maximum (FWHM) of the nonlinear signal electric field obtained from a Gaussian function fit of the time and time delay dependent amplitude for DFWM and magic angle DFWM.

\section{Conclusion}
In conclusion, we have performed complete electric field measurement of femtosecond time-resolved four-wave mixing signals in a DFWM scheme from carbon dioxide molecules. Using the ultrashort, ultra-weak pulse measurement technique of TADPOLE, we have measured the temporal electric field of pico-joule level DFWM signals as a function of time delay between the interacting DFWM pulses.  These measurements were performed in the standard DFWM configuration as well as in the magic angle DFWM configuration where the rotational contribution is suppressed. The chirp of the nonlinear signal electric field is found to sensitively depend on the electronic contribution which enhances the time delay-dependent chirp compared to the chirp of the input probe pulse. No enhancement of the chirp is observed for the rotational contribution. The electric field measurement thus allows us to extract the electronic nonlinear response even when the rotational contribution dominates the overall nonlinear signal. Our work demonstrates that ultrafast electric field-resolved nonlinear spectroscopy can be a sensitive measure of electron dynamics. When applied to nonlinear spectroscopy of excited states \cite{marek2011,thurston2020jpca}, such as molecules and materials excited by attosecond pulses, electric field measurements can track the induced transient nonlinear polarization, which could offer new insight into ultrafast coherent electron dynamics.  

%In this work, we have shown a method for measuring the emitted electric field from a nonlinear optical process, and applied it to study third-order response in CO$_{2}$. We have studied the emitted electric field as a function of gate-probe time delay. Since the emitted electric field contains the total imprint of the involved nonlinear process, higher order nonlinear processes, such as those dependent on $\chi^{5}$, can possibly also be probed with additional modelling \cite{ekvall2001}. The chirp of the emitted signal field is studied as a function of gate-probe delay.

\begin{backmatter}
% \bmsection{Funding}
%  Content in the funding section will be generated entirely from details submitted to Prism. Authors may add placeholder text in the manuscript to assess length, but any text added to this section in the manuscript will be replaced during production and will display official funder names along with any grant numbers provided. If additional details about a funder are required, they may be added to the Acknowledgments, even if this duplicates information in the funding section. See the example below in Acknowledgements.

\bmsection{Acknowledgments}
Work at the Molecular Foundry was supported by the Office of Science, Office of Basic Energy Sciences, of the U.S. Department of Energy under Contract No. DE-AC02-05CH11231.

\bmsection{Disclosures}
The authors declare no conflicts of interest.

\bmsection{Data availability} Data underlying the results presented in this paper are not publicly available at this time but may be obtained from the authors upon reasonable request.

\end{backmatter}

%%%%%%%%%%%%%%%%%%%%%%% References %%%%%%%%%%%%%%%%%%%%%%%%
%%%%%%%%%% If using BibTeX:
\bibliography{references}

\end{document}

\maketitle

\section{TADPOLE Electric Field Retrieval}
\begin{figure}[htbp]
\centering\includegraphics[width=9cm]{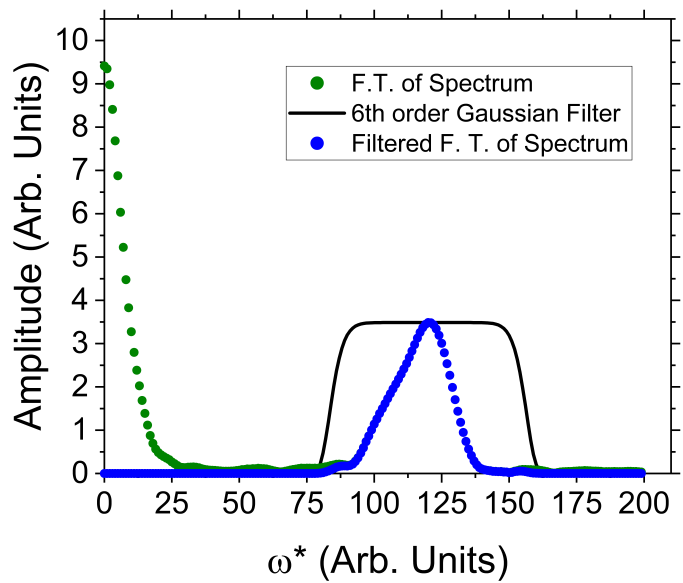}
\caption{Fourier transform (F.T.) of the spectrum of the Signal and Reference beams combined. The peak corresponding to fast oscillations is filtered using a 6th order Gaussian filter and shifted to zero to remove the linear phase term $\omega\tau_{R}$.}
\label{figS1:FT_Spectrum}
\end{figure}

Fig. \ref{figS1:FT_Spectrum} shows the Fourier transform of a combined signal and reference spectra.

The cosine term in equation 1 in the main text is the fast oscillating term and the signal phase is contained in the high-frequency peak of the Fourier transform. There is a linear phase from the $\omega \tau_{R}$ term that is removed by shifting the peak to 0 in the $\omega^*$ domain. Inverse Fourier transforming back to the $\omega$ domain allows access to the signal phase. The phase extracted from the inverse Fourier transform also contains the reference phase which is subtracted, to retrieve the signal phase.

\begin{figure}[t]
\centering\includegraphics[width=12cm]{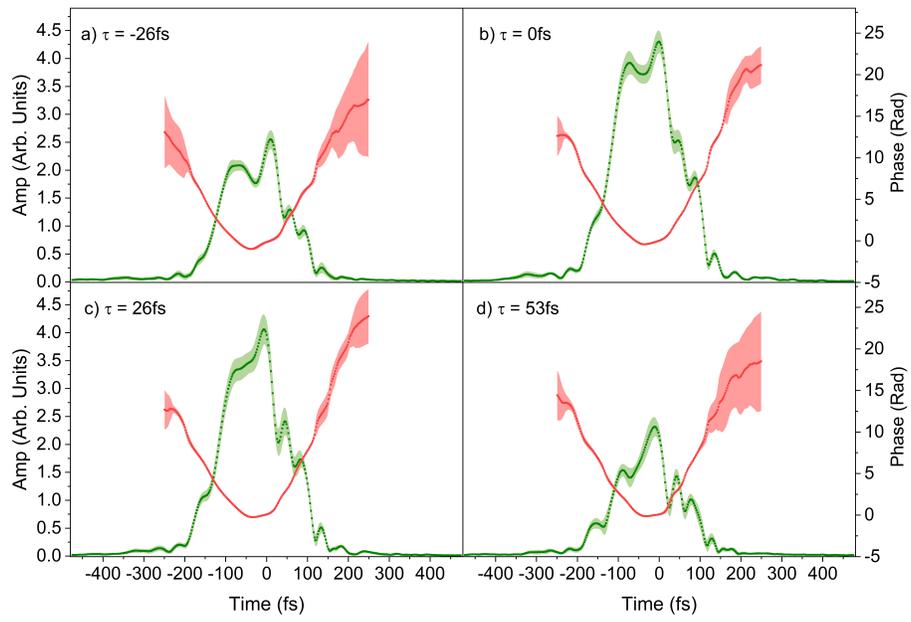}
\caption{Measured Signal electric field $E(t)$ for magic angle DFWM, for Gate-Probe time delays (a) -26.6 fs, (b) 0 fs, (c) 26.6fs fs, and (d) 53 fs. Electric field amplitude $E_0(t)$ and phase $\phi(t)$ for magic angle DFWM are in green and red, respectively.}
\label{figS2:E_t_timedelay}
\end{figure}

Figure \ref{figS2:E_t_timedelay} shows the lineouts of signal electric field amplitude and phase from figure 3 in the main text, at four different Gate-Probe time delays $\tau$.

% Fig. \ref{figS2:DFWM_Ortho_Multiple_Intensities} shows the DFWM signal electric field for the case when the analyzer is orthogonal to the probe polarization. Multiple intensities of the gate are show to demonstrate the lack of plasma contribution.At these low intensities we do not expect to see any plasma in carbon dioxide, additionally we do not see any significant changes to the chirp or amplitude fit parameters. 
% \begin{figure}[htbp]
% \centering\includegraphics[width=10cm]{Sup._Docs/FigS3_Amp_bparam_PNG.png}
% \caption{Magnitude of the measured E-field and chirp as a function of delay between the gate and probe pulses for the case of analyzer angle orthogonal to probe polarization for multiple gate intensities.}
% \label{figS2:DFWM_Ortho_Multiple_Intensities}
% \end{figure}

% Bibliography
%\bibliography{references}